\documentclass[aps,prl,twocolumn,groupedaddress]{revtex4}

\usepackage{graphicx}

\begin{document}

\title{Layered tungsten oxide-based organic/inorganic hybrid materials \\ I: Infrared and Raman study}

\author{B. Ingham}\affiliation{Victoria University of Wellington, P.O. Box 600,
Wellington, New Zealand}

\author{S.V. Chong} \author{J.L. Tallon} \affiliation{Industrial Research Ltd.,
P.O. Box 31310, Lower Hutt, New Zealand.}

\date{\today}

\begin{abstract}
Tungsten oxide-organic layered hybrid materials have been studied
by infrared and Raman spectroscopy, and demonstrate a difference
in bonding nature as the length of the interlayer organic `spacer'
molecule is increased. Ethylenediamine-tungsten oxide clearly
displays a lack of terminal $-$NH$_3$$^+$ ammonium groups which
appear in hybrids with longer alkane molecules, thus indicating
that the longer chains are bound by electrostatic interactions as
well as or in place of the hydrogen bonding that must be present
in the shorter chain ethylenediamine hybrids. The presence of
organic molecules between the tungsten oxide layers, compared with
the layered tungstic acid H$_2$WO$_4$, shows a decrease in the
apical W=O bond strength, as might be expected from the
aforementioned electrostatic interaction.
\end{abstract}

\maketitle

\section{Introduction}

Layered transition metal oxides present a paradigm for strongly
correlated electronic systems subject to interesting correlated
states and the competition of multiple order parameters. The high
temperature oxide superconductors are a much-studied case in
point. Moreover the discovery of superconductivity in sodium
cobalt oxide \cite{NaCoO} confirms this general approach of
exploring novel layered metal oxides. Here we consider layered
tungstates as another new model system capable of displaying a
complex phase behaviour dependent upon interlayer coupling and
doping state.

Tungsten oxide has been the subject of much research in the past
few decades due to its interesting physical and electronic
properties and doping capability. Tungsten trioxide can form a
variety of stable structures at room temperature, such as
pyrochlore, hexagonal, or distorted cubic
\cite{Mourey,Guo,Genin,Hagenmuller}, all of which are comprised of
corner- and/or edge-shared WO$_6$ octahedra. These structures can
be doped by inserting mono- or di-valent atomic species into the
interstitial vacancy sites within the oxide structure
\cite{Hagenmuller,Rollinson,Haydon,Grenthe}, or by removing oxygen
\cite{Booth,Georg,Glemser}. This results in an increase in
electrical conductivity \cite{Glemsauer,Sienko,Brown,Gardner} and
dramatic colour changes \cite{Georg,Glemsauer,Straumanis}, which
has led to tungsten trioxide systems being used in electrochromic
applications \cite{Ahn,Goldner,Bange}.

Tungsten trioxide can be hydrated to form layered structures, with
a general formula of WO$_3$$\cdot$$\textsl{x}$H$_2$O. In the
mono-hydrate, WO$_3$$\cdot$H$_2$O (or H$_2$WO$_4$),
two-dimensional layers of corner-shared WO$_6$ octahedra are
formed, with a water molecule ($-$OH$_2$) substituted for one
apical oxygen of the tungsten atom, and a terminal oxygen
completing the structure \cite{Szymanski}. In the di-hydrate
(WO$_3$$\cdot$2H$_2$O), the second water molecule is inserted
between the layers, as in MoO$_3$$\cdot$2H$_2$O \cite{Krebs}. An
increase in the interlayer spacing is observed \cite{Daniel} and
it thus follows that one could perhaps substitute organic species
between the tungsten oxide layers. Indeed this has proven
successful, with various groups succeeding in intercalating
organic molecules such as pyrazine \cite{Yan}, 4,4-bipyridine
\cite{Yan,Johnson}, pyridine \cite{Johnson}, tert-alkylammonium
species \cite{Kikkawa}, mono-aminoalkanes \cite{Ayyappan} and
diamino-alkanes \cite{Chong}. Such materials provide an
interesting template for doping and substitutional (e.g. magnetic
ions) studies in low-dimensional systems.

Here we present infrared and Raman results on a selection of
WO$_3$-based hybrid materials in order to gain a more extensive
understanding of the overall structure; in particular, the bonding
nature of the organic to the inorganic layer, and the impact of
the organic intercalant on the structure of the inorganic layer.

\section{Experimental}

Diaminoalkane-tungsten oxide samples were synthesised as described
in Ref. \onlinecite{Ingham}. In brief, H$_2$WO$_4$ was dissolved
in hot aqueous ammonia solution (33 wt. $\%$) and then a solution
of the organic amines ($\alpha$,$\omega$-diamine,
H$_2$N(CH$_2$)$_n$NH$_2$, hereby abbreviated to DAn; or
phenethylamine, C$_6$H$_5$(CH$_2$)$_2$NH$_2$, abbreviated to phen)
dissolved in ammonia was added in a 2:1 molar ratio. The excess
solvent was then evaporated off and the hybrid materials were
obtained as white/cream-coloured powders. The entire synthesis was
carried out under flowing N$_2$ gas.

Powder X-ray diffraction spectra were recorded using a Philips
PW1700 series powder diffractometer with Co K$\alpha$ radiation.
Infrared spectra were collected on a Bomem DA8 FT spectrometer
over the mid-IR range (450-4000 cm$^{-1}$) using the KBr disc
method with a resolution of 2 cm$^{-1}$. Raman spectra were
collected using a Jobin-Yvon LabRam HR spectrometer with an
excitation wavelength of 633 nm.

\section{Results and Discussion}

\begin{figure}
\includegraphics*[height=110mm, width=85mm]{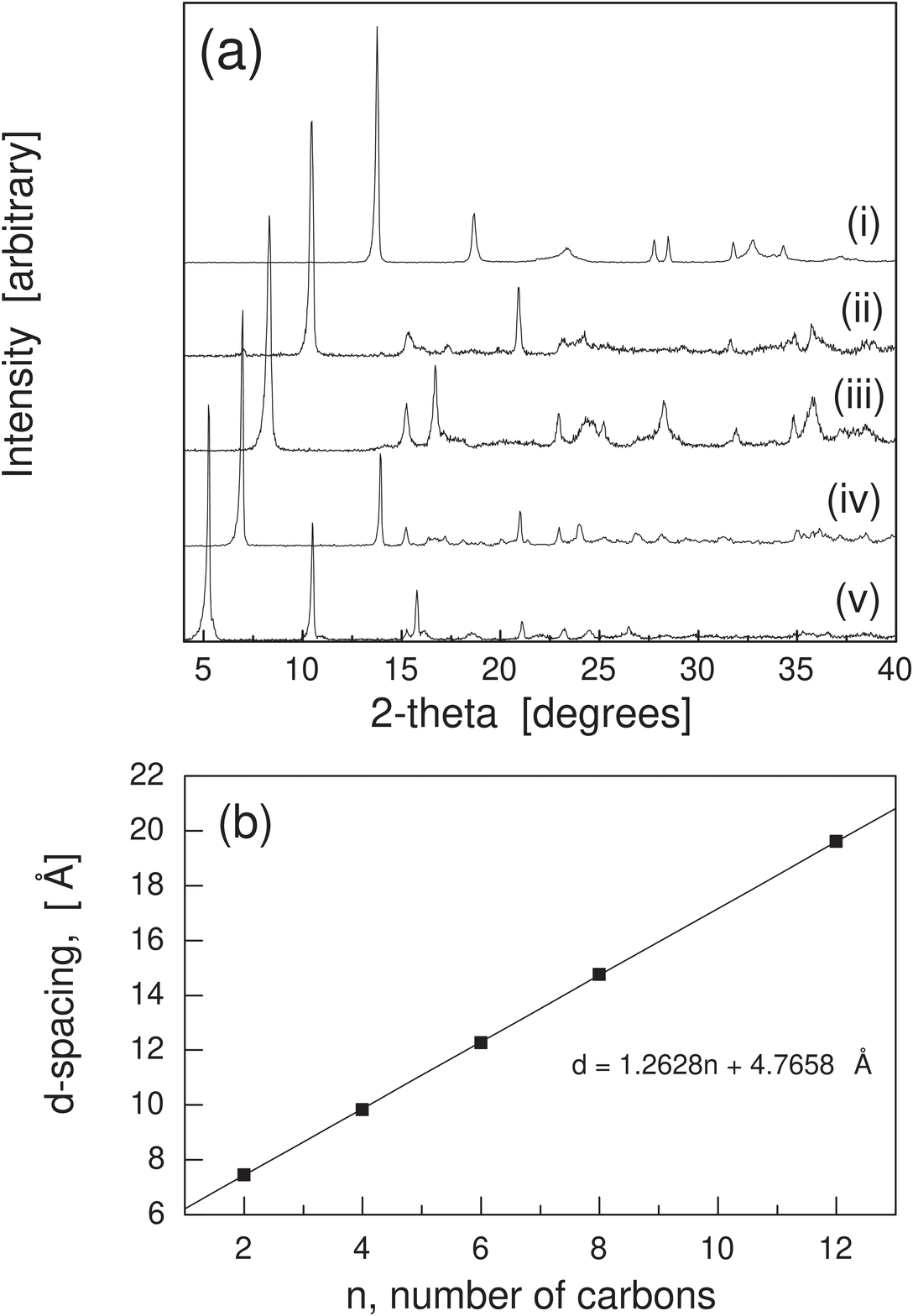}
\caption{\label{fig:xrd} (a) X-ray diffraction spectra of tungsten
oxide-based hybrid materials: (i) W-DA2; (ii) W-DA4; (iii) W-DA6;
(iv) W-DA8; (v) W-DA12. (b) Linear progression of first major peak
versus alkane chain length.}
\end{figure}

The X-ray diffraction spectra shown in Figure~\ref{fig:xrd}a
display a strong peak at low 2$\theta$ angles. As the length of
the intercalated organic molecule is increased, the corresponding
d-spacing also increases in a linear fashion (shown in
Figure~\ref{fig:xrd}b). From the slope and intercept of this line
we propose a model by which single molecular layers of
2-dimensional corner-shared WO$_6$ octahedra are separated by the
organic molecules which are aligned almost perpendicular to the
tungsten oxide layers. One would expect such a layered compound to
display properties approaching those of a purely 2-dimensional
system as the distance between the oxide layers is increased (by
increasing the length of the organic spacer molecule). The
physical properties of this system will be the topic of future
publications.

\begin{figure}
\includegraphics*[height=60mm, width=85mm]{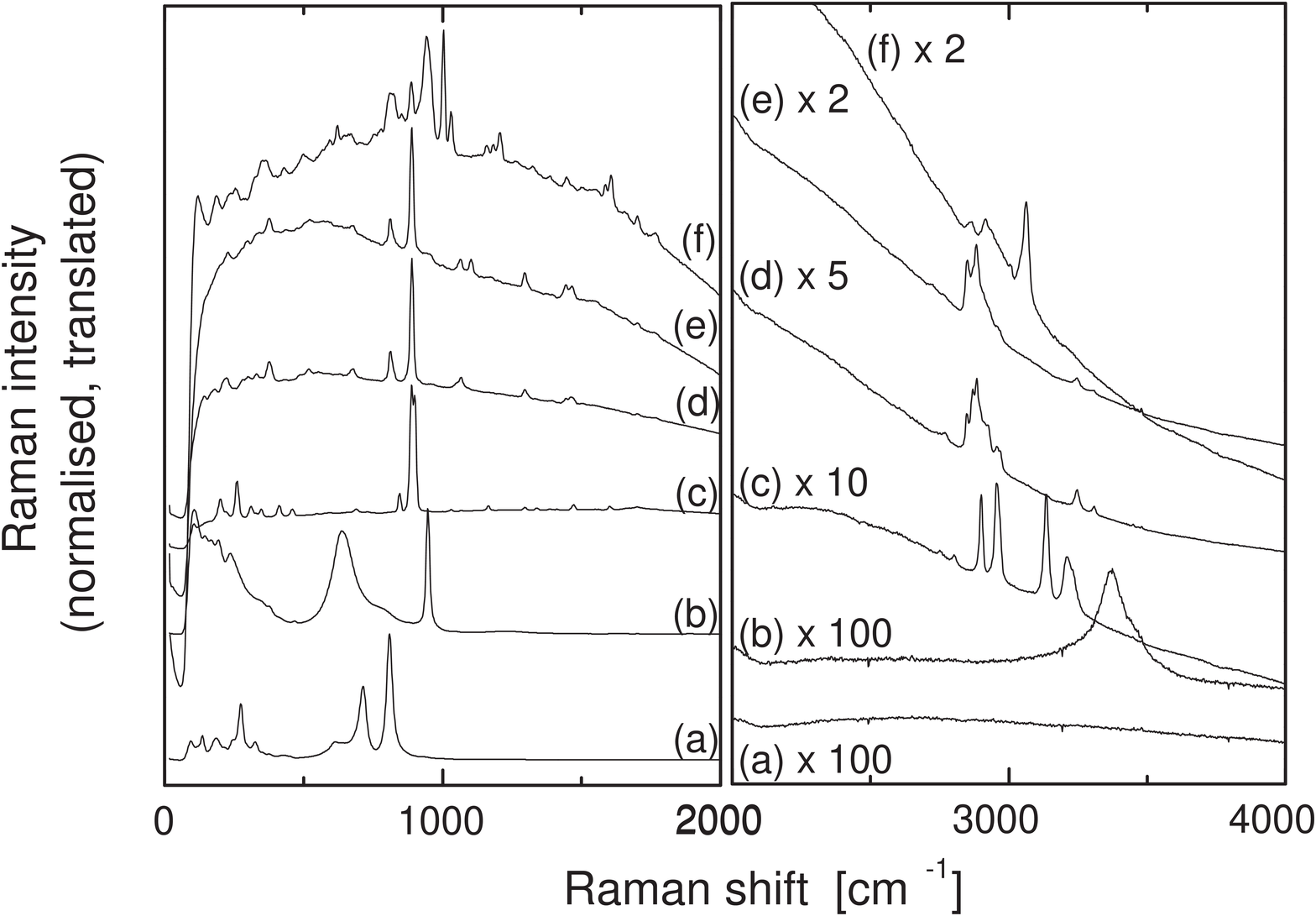}
\caption{\small \label{fig:Raman} Raman spectra of tungsten
oxide-related materials: (a) WO$_3$; (b) H$_2$WO$_4$; (c) W-DA2;
(d) W-DA6; (e) W-DA12; (f) W-phen. The high frequency region is
displayed on a different scale to enable the important features to
be seen.}
\end{figure}

\begin{figure}
\includegraphics*[height=60mm, width=85mm]{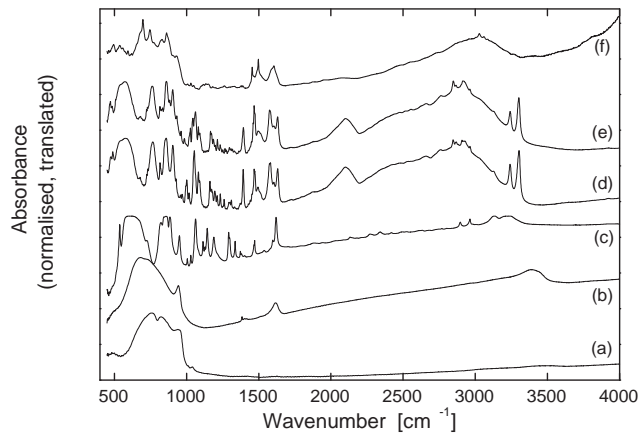}
\caption{\label{fig:IRfull} Infrared spectra of tungsten
oxide-related materials: (a) WO$_3$; (b) H$_2$WO$_4$; (c) W-DA2;
(d) W-DA6; (e) W-DA12; (f) W-phen.}
\end{figure}

\begin{figure}
\includegraphics*[height=60mm, width=85mm]{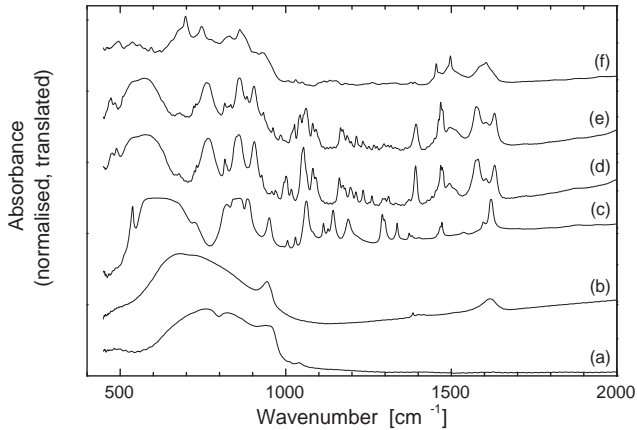}
\caption{\small \label{fig:IRzoom} Low-frequency region of the
infrared spectra of tungsten oxide-related materials: (a) WO$_3$;
(b) H$_2$WO$_4$; (c) W-DA2; (d) W-DA6; (e) W-DA12; (f) W-phen.}
\end{figure}

The normalised infrared and Raman spectra of the solid samples
studied are shown in Figures \ref{fig:Raman}-\ref{fig:IRzoom}, and
the peak positions tabulated in Tables \ref{table1}-\ref{table2}.
The diaminoalkane hybrids WO$_4$$\cdot$DAn (hereby abbreviated to
W-DAn) with alkyl lengths longer than two carbons have virtually
identical spectra \cite{Chong,Ingham} and so are summarised in the
tables as W-DAn, in comparison with the differing W-DA2.

The infrared spectra of WO$_3$ and its hydrate, H$_2$WO$_4$, have
much fewer and broader peaks than their hybrid counterparts. Both
the infrared and Raman spectra of these two compounds correspond
well with those presented by Daniel et al. \cite{Daniel}. These
results are summarised in three sections: the effect on the
inorganic layer, the effect on the organic species, and the
organic-inorganic bonding nature.

\begin{table*}
\caption{\label{table1}Observed peaks in powder samples, 1000-4000
cm$^{-1}$. $\nu$ stretching, $\delta$ deformation/in-plane
bending, $sciss.$ scissoring, $\sigma$ bending, $\omega$ wagging;
$s$ strong, $m$ medium, $w$ weak, $vw$ very weak, $b$ broad.}
\begin{ruledtabular}
\begin{tabular}{lcccccccccc}
Assignment&
\multicolumn{2}{c}{WO$_3$}&\multicolumn{2}{c}{H$_2$WO$_4$}&\multicolumn{2}{c}{W-DA2}&\multicolumn{2}{c}{W-DAn}&\multicolumn{2}{c}{W-phen}\\
&IR&Raman&IR&Raman&IR&Raman&IR&Raman&IR&Raman\\
\hline $\nu$(O$-$H)&&&3410 mb&3370 w\\
\hline &&&&&&&3302 s&3305 vw\\
&&&&&3244 wb&&3242 m&3247 w\\
&&&&&3212 wb&3212 m&\\
$\nu$(NH$_2$,NH$_3$$^+$)&&&&&3134 wb&3135 s\\
&&&&&&&&&3061 w&3063 m\\
&&&&&&&&&3028 w\\
&&&&&&&&&3006 w\\
\hline &&&&&2963 w&2956 s&2963 w&2960 w\\
&&&&&2895 w&2900 s&2919 w&2924 w&&2916 w\\
$\nu$(CH$_2$)&&&&&&&2910 w&2883 m\\
&&&&&&&2868 w&2870 m&&2861 w\\
&&&&&&&2847 w&2849w\\
\hline $\delta$(NH$_3$$^+$)&&&&&&&2101 mb&&2090 vw\\
\hline Organic&&&&&&1762 vw&&1760 vw&&1766 w\\
&&&&&&1700 vw&&1701 vw&&1702 w\\
\hline $\delta$(H$_2$O)&&&1614 m\\
\hline &&&&&1620 s&&1631 s\\
$\delta$(NH$_2$/NH$_3$$^+$)&&&&&1597 w&1602 vw&1604 w&&1605 mb&1606 w\\
&&&&&&&1578 s&&1591 mb&1586 w\\
\hline Sciss. CH$_2$&&&&&1471 m&1472 w&1495 m&&1497\footnotemark[1] m&1500\footnotemark[1] vw\\
&&&&&1467 w&&1469 s&1465 w\\
&&&&&&&1445 vw&1444 w&1454\footnotemark[1] m&1446\footnotemark[1] w\\
Sciss. and $\sigma$(CH$_2$)&&&&&1403 vw&1412 vw\\
&&&&&1382 vw&1386 vw&1393 s&&1384
vw&1386 vw\\
&&&&&1373 w\\
$\sigma$(CH$_2$), $\omega$(CH$_2$)&&&&&1336 m&1340 vw&1310 w&&1333
vw&1322 vw\\
&&&&&1299 m&1296 vw&1298 w&1296 w\\
\hline &&&&&1292 m&&1263 w&&1261
vw&1266 vw\\
&&&&&&&1213 w&&&1206 m\\
&&&&&1189 m&&1197 w\\
$\sigma$(NH$_2$, NH$_3$$^+$)&&&&&&&1173 m&&1181 vw&1183 w\\
&&&&&1144 m&1164 vw&1164 m&&1152 vw&1161 w\\
&&&&&1128 vw&&&&1132 vw\\
&&&&&1114w\\
\hline &&&&&&&&1103 w\\
&&&&&1062 s&1070 vw&1083 m&1066 w\\
$\nu$(C$-$N, C$-$C)&&&&&&&1059 s\\
&&&&&1030 w&1030 vw&1027 m&&&1030 m\\
&&&&&1006 w&&1001 m\\
&&&&&&&962 w&953 vw&929 mb&942 sb\\
\end{tabular}
\end{ruledtabular}
\footnotetext[1]{Aromatic C=C stretch.}
\end{table*}

\begin{table*}
\caption{\label{table2}Observed peaks in powder samples, 0-1000
cm$^{-1}$. $\nu$ stretching, $\delta$ deformation/in-plane
bending, $\rho$ rocking; $s$ strong, $m$ medium, $w$ weak, $vw$
very weak, $b$ broad.}
\begin{ruledtabular}
\begin{tabular}{lcccccccccc}
Assignment&
\multicolumn{2}{c}{WO$_3$}&\multicolumn{2}{c}{H$_2$WO$_4$}&\multicolumn{2}{c}{W-DA2}&\multicolumn{2}{c}{W-DAn}&\multicolumn{2}{c}{W-phen}\\
&IR&Raman&IR&Raman&IR&Raman&IR&Raman&IR&Raman\\
\hline $\nu$(W=O)&940 mb&940 vw&950 s&947 s&951? m&900 s&&&&1004?
s\\
&&&&&885 s&888 s&905 s&890 s&902 wb&888 m\\
\hline &&&&&857 sb&&859 s&&861 s&854 m\\
&&&&&841 sb&846 m\\
&824 sb&809 s&&&822 mb&&816 m&813 m&829 m&816 mb\\
$\nu$(O$-$W$-$O)&756 sb&&&774 mb&&&760\footnotemark[2]
sb&&775\footnotemark[2] mb\\
&&714 m&740 sb&&724 mb&732 vw&&&745 s\\
&&&680 sb&&674 sb&690 w&678 vw&679 w&697 s&668 w\\
&&614 w&&638 sb&613 sb&&&&&622 m\\
&&&&&&594 vw&577 sb&576 vw&594 vw&595 w\\
\hline &&&&&&550 vw&545 sb&552 vw&557 w\\
Organic&&&&&537 s&&&519 w&538 w&502 m\\
&&&&&&&487 w&490 vw&495 w\\
&&&&&&460 w&474 w\\
\hline Inorganic&&434 vw&&467 vw&&412 w&&436 vw&&432 w\\
\hline W$-$OH$_2$&&&&377 vw&&&&377 m&&362 mb\\
\hline &&326 w&&330 vw&&348 w&&336 vw\\
$\delta$(O$-$W$-$O)&&&&&&311 w&&301 vw\\
&&274 m&&&&261 m&&283 vw\\
\hline $\nu$(W$-$O$-$W)&&252 w&&236 w&&222 w&&226 w&&254 w\\
&&&&&&&&&&238 w\\
\hline &&187 w&&194 w&&201 m&&182 vw&&187 w\\
Lattice modes&&136 w&&144 vw&&&&144 vw&&120 vw\\
&&94 w&&106 w\\
\end{tabular}
\end{ruledtabular}
\footnotetext[2]{$\rho$(NH$_3$$^+$) mode.}
\end{table*}

\subsection{I. The effect on the inorganic layer}

The presence of co-ordinated water molecules in H$_2$WO$_4$ can be
seen by the broad O$-$H stretching peak at 3410 (3370) cm$^{-1}$
(IR and Raman respectively), the H$_2$O bending peak at 1614
cm$^{-1}$ (IR only), and the W$-$OH$_2$ co-ordinated water peak in
the Raman spectrum at 377 cm$^{-1}$. These peaks do not occur in
the WO$_3$ spectra, as expected.

As mentioned earlier, the structure of H$_2$WO$_4$ consists of
layers of corner-shared WO$_6$ octahedra with alternate apical
arrangements of W$-$OH$_2$ and W=O. In the Raman spectrum the W=O
bonding is shown clearly by a sharp band centred at about 950
cm$^{-1}$. This is also present in the IR spectrum. Surprisingly,
the WO$_3$ sample also exhibits a small peak at this position in
the Raman and IR spectra. This is not expected as the structure of
WO$_3$ consists only of single W$-$O bonds. However it can be
explained by the presence of disorder in the sample and loss of
oxygen \cite{Lee}\footnote{Commercial WO$_3$ powder is a pale
yellow colour, which is reduced over 1-2 days in air to a pale
green colour. This is indicative of a loss of oxygen but can be
regained by storing in an oxygen atmosphere or heating in oxygen
for a few hours.}, which results in the formation of a small
fraction of W=O bonds. There are also W=O terminations on the
surfaces of the powder particles.

In the diaminoalkane hybrid compounds the characteristic W=O peak,
formerly at 950 cm$^{-1}$ in H$_2$WO$_4$, shifts to a lower
frequency of 890-900 cm$^{-1}$. While in several of these
compounds there are peaks at 950 cm$^{-1}$ they are not as intense
as the W=O peak. It is also of interest to note that the peak at
888 cm$^{-1}$ in the W-DA2 Raman spectrum is a doublet, suggesting
perhaps the presence of two different W=O bonds. In W-phen, the
peak structure in the range 900-100 cm$^{-1}$ is quite different
from the other hybrid spectra. This may be due to the organic
molecule being mono-dentate (as IR spectra of other
mono-aminoalkane hybrids show that the 950 cm$^{-1}$ W=O peak is
virtually unaltered \cite{Chong2}) or more complicated due to the
presence of the aromatic ring. In any case, the sharp Raman peak
at 1004 cm$^{-1}$ is likely to be a W=O peak due to the similarity
in shape between it and the known W=O peak in H$_2$WO$_4$, for
example. There may be more than one variation of the W=O bond
within the structure, which may also account for the multiple
peaks observed in this region, as is the case in
WO$_3$$\cdot$2H$_2$O \cite{Daniel}. There is a relationship
between bond length (an indication of bond strength) and the
frequency of the vibrational modes(s) for the terminal W$-$O bond,
namely that as the bond length decreases, bond strength and
frequency increases \cite{Daniel,Cotton}. It is not therefore
impossible to observe more than one terminal W$-$O vibration.

The remaining bands can be assigned as follows: 580-860 cm$^{-1}$
O$-$W$-$O stretching, 430-470 cm$^{-1}$ inorganic Raman mode
(W$-$O), 260-350 cm$^{-1}$ O$-$W$-$O bending, 220-250 cm$^{-1}$
W$-$O$-$W stretching, 90-200 cm$^{-1}$ lattice modes. These bands
occur at similar positions in the hybrid compounds.

\subsection{II. The effect on the organic species}

Perhaps the most telling difference between W-DA2 and W-DAn
(n$>$2) is the presence or absence of the broad peak centred at
~2100 cm$^{-1}$ in the IR spectra. This feature is due to a
combination of the asymmetrical $-$NH$_3$$^+$ bending vibration
and the torsional oscillation of the $-$NH$_3$$^+$ group
\cite{Silverstein}. Both $-$NH$_2$ scissoring and $-$NH$_3$$^+$
bending occurs in the region 1580-1630 cm$^{-1}$, however the
latter displays three peaks (as seen in W-DAn) while the former
yields only two (as in W-DA2). Lastly, the N$-$H stretching bands
occur at slightly higher energies in W-DAn compared with W-DA2
(3240-3300 cm$^{-1}$, c.f. 3135-3250 cm$^{-1}$) and these peaks
are also much sharper. However, an increase in alkyl chain length
should have little effect on the N$-$H stretching frequency
\cite{Segal2}. The observed differences are discussed in Section
III.

The presence of broad vibrational bands in the W-DAn spectra in
the range 2430-2790 cm$^{-1}$ can be ascribed to NH$_3$$^+$
symmetrical stretching \cite{Silverstein,Segal1} and often appear
as a broad, ill-defined band under the C$-$H stretching modes near
2800 cm$^{-1}$ \cite{Baldwin}. In H$_2$WO$_4$ a small peak is
observed in the IR spectrum at 1614 cm$^{-1}$, which corresponds
to H$_2$O bending. While the hybrid samples also exhibit peaks
near this value (1600-1630 cm$^{-1}$), these peaks are sharp and
well defined, and the vibrational band at ~3400 cm$^{-1}$
(indicative of O$-$H stretching) is absent. Thus we conclude that
most of the co-ordinated water molecules (W$-$OH$_2$) have been
dehydrogenated during the organic intercalation process (a small
feature at 377 cm$^{-1}$ in the Raman spectra may correspond to
W$-$O$-$H). The peaks in the hybrid spectra in the range 1600-1630
cm$^{-1}$ correspond instead to NH$_2$ or NH$_3$$^+$ bending.

As only the terminal amino/ammonium group is involved in
interactions with the inorganic layer, it is no surprise that the
bands corresponding to C$-$H, C$-$N and C$-$C vibrations remain
virtually unchanged. The peaks in the hybrid spectra corresponding
to these vibrations can be assigned as follows: 2850-2960
cm$^{-1}$ C$-$H stretching, 1300-1500 cm$^{-1}$ CH$_2$ wagging,
bending, and scissoring; also C=C aromatic stretching for W-phen
(1450-1500 cm$^{-1}$), 950-1100 cm$^{-1}$ C$-$N and C$-$C
stretching. From 460-550 cm$^{-1}$ there are a small number of
unassigned organic bands (including a sharp peak in the IR at 537
cm$^{-1}$ for W-DA2, corresponding to the bending vibration of the
NCCN backbone \cite{Sabatini}).

\subsection{III. The organic-inorganic bonding nature}
The most noticeable difference among the hybrid samples is that
W-DA2 shows quite a different bonding nature to the hybrids with
longer organic chains: there is no evidence for the presence of
ammonium ($-$NH$_3$$^+$) terminal groups in W-DA2. As mentioned
previously, W-DA2 lacks a feature at 2100 cm$^{-1}$ found in the
other hybrid spectra that corresponds to deformations of terminal
$-$NH$_3$$^+$. Combined with the presence of the $-$NH$_3$$^+$
rocking mode found at ~770 cm$^{-1}$ \cite{Segal1}, this implies
that some of the amine molecules in W-DAn appear as
R$-$NH$_3$$^+$$\cdots$$^-$O$-$W, while all those of W-DA2 appear
as neutral $-$NH$_2$ species. This raises the question of how
charge balance is achieved in W-DA2 and also whether hydrogen
bonding is the exclusive mechanism of bonding, whereas in the
longer organic molecule hybrids there is clearly an electrostatic
component as well due to the presence of terminal ammonium groups.
There are also differences in the N$-$H stretch region, suggesting
changes in the bonding chemistry as one progresses from DA2 to
longer DAn alkyl chains. As mentioned in Section II, the N$-$H
stretching bands for W-DA2 appear at lower wavenumbers than for
longer-chain W-DAn species. Hydrogen bonding (which holds the DA2
molecules in place between the layers) will cause these bands to
shift to lower frequencies \cite{Silverstein}. DA2 is the only
case where the inductive effect on the terminal amine groups is
known to be a perturbing factor to the molecular vibrations due to
the shorter alkyl chain length \cite{Segal2}.

\begin{figure}
\includegraphics*[height=60mm, width=85mm]{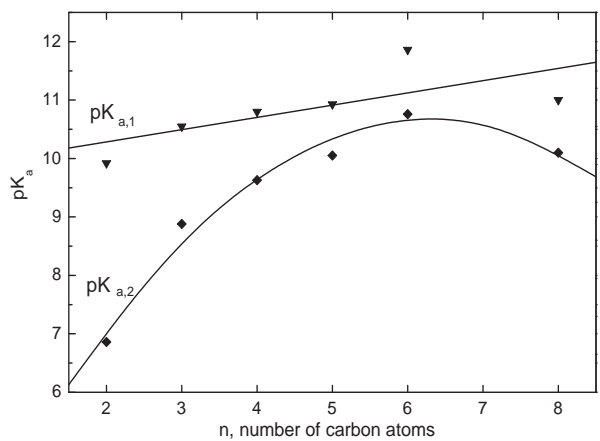}
\caption{\small \label{fig:pka} Plot of pK$_a$ values from Ref.
\onlinecite{CRC} versus the number of carbon atoms of
$\alpha$,$\omega$-diamines in aqueous solution.}
\end{figure}

One possible explanation for the observed difference in structure
between DA2 and longer DAn molecules intercalated into an acidic
metal oxide, arises from examining the trend in basicity of
$\alpha$,$\omega$-diamines in solution. Although the trend in
basicity only changes slightly with increasing alkyl length for
the monoamines (due to combinations of solution effect, steric
hindrance to solvation, and number of available H-bonds
\cite{Brauman,Arnett,Aue}), a substantial increase in basicity is
observed in going from DA2 to DA8 (Fig.~\ref{fig:pka}). This
increase in base strength for the diaminoalkanes is due to an
increase of the inductive effect as the number of methylene groups
($-$CH$_2$$-$) increases. The formation of alkylammonium
($-$NH$_3$$^+$) end groups will therefore be more favourable for
the more basic diaminoalkanes (i.e. diaminobutane and above). This
apparent difference is also manifested in the TGA profile of these
hybrids, with W-DA2 dissociating at a higher temperature than any
of the other W-DAn compounds \cite{Chong}.

The strong Raman peak at 950 cm$^{-1}$ in H$_2$WO$_4$ corresponds
to an apical W=O bond, which is characteristic of a layered
structure. This peak is present in all of the hybrid compounds
although it is shifted to lower wavenumbers of 890-900 cm$^{-1}$,
indicating that the apical oxygen is not as tightly bound to the
tungsten as in H$_2$WO$_4$. This too can be expected as charge
balance in H$_2$WO$_4$ is achieved by co-ordinated water molecules
binding alternately upper and lower of the tungsten layer, leaving
the remaining apical oxygen atoms to bind relatively strongly. In
the hybrid materials one might expect the influence of organic
molecules on each apical oxygen to be more uniform, hence this
interaction will weaken the W=O bond, and this is indeed evident
from the Raman shift. There is evidence to suggest that in W-phen,
a monodentate amine hybrid, there are several variations of the
W=O bond, which in itself warrants further study.

The structure and electronic density of states of these layered
systems are currently being investigated by $\textit{ab initio}$
computations. The question of the relocation of protons in the
W-DA2 system is a particular issue of study.

\section{Conclusions}

Infrared and Raman spectroscopy have been utilised to gain an
understanding of the bonding nature between the organic and
inorganic components of tungsten oxide-hybrid systems with
different length diaminoalkanes and how the inclusion of organic
molecules affects the structure of the inorganic layers. Tungsten
oxide-ethylenediamine displays a different bonding nature from the
other hybrid materials in that it appears to rely on
hydrogen-bonding only, with no terminal $-$NH$_3$$^+$ ammonium
groups to form electrostatic bonds. This is thought to be related
to the lower basicity of ethylenediamine as opposed to the longer
diaminoalkane species. All of the tungsten oxide-organic hybrid
materials exhibit a decrease in the strength of the apical W=O
bond of the WO$_4$$^{2-}$ layers compared with the layered
tungsten oxide monohydrate (tungstic acid), H$_2$WO$_4$.

\section{Acknowledgments}

The authors would like to acknowledge the financial assistance
from The New Zealand Foundation of Research Science and Technology
(Contract number: IRLX0201), The Royal Society of New Zealand
Marsden Fund, and The MacDiarmid Institute for Advanced Materials
and Nanotechnology (Victoria University, New Zealand).

\end{document}